# Influence of the Dzyaloshinskii-Moriya interaction on the spin-torque diode effect

R. Tomasello,[1] M. Carpentieri,[2] G. Finocchio[3]

[1]Department of Computer Science, Modelling, Electronics and System Science, University of Calabria, Rende (CS), Italy.

[2]Department of Electrical and Information Engineering, Politecnico of Bari, via E. Orabona 4, I-70125 Bari, Italy.

[3]Department of Electronic Engineering, Industrial Chemistry and Engineering, University of Messina, C.da di Dio, I-98166, Messina, Italy.

**Abstract:** This paper predicts the effect of the Dzyaloshinskii-Moriya interaction (DMI) and spin Hall effect in the spin-torque diode response of a Magnetic Tunnel Junction built over a Tantalum strip. Our results indicate that, for a microwave current large enough, the DMI can change qualitatively the resonant response by splitting the ferromagnetic resonance peak. We also find out that the two modes have a non-uniform spatial distribution.

MgO-based magnetic tunnel junctions (MTJs) are attracting much interest for their several technological applications, such as spin transfer torque magnetic random access memories (STT-MRAM) and nano-oscillators[1, 2, 3, 4, 5]. However, when a microwave current $J_{MTJrf}$ flows through an MTJ with a frequency close to the ferromagnetic resonance (FMR) frequency of the free-layer (FL), the so called spin-torque diode effect can be observed.[6, 7] In other words, the tunnelling magnetoresistance (TMR) oscillates at the same frequency of the microwave current and, as a result, a dc voltage can be measured across the MTJ stack. The spin diode effect can be also used to estimate the STT-field-like torque term and its voltage dependence.[8, 9] In the MTJ configuration where the FL is coupled to a heavy metal, we can use the additional degree of freedom regarding the spin-Hall effect (SHE)[10] (the electrical current is converted into a transversal spin current $J_{SHE}$) in controlling the spin-torque diode response. When an ultra-thin ferromagnetic layer is coupled to a heavy metal with strong spin-orbit coupling, because of the interfacial Dzyaloshinskii-Moriya interaction (DMI), a chiral magnetic field should be also taken into account.[11, 12, 13, 14, 15] The DMI is an antisymmetric interfacial exchange contribution due to the spin-orbit coupling that excites spatial rotational magnetization configurations, such as spirals, skyrmions and chiral structures.[16] The DMI is relevant in bulk non-centrosymmetric crystal lattice, and in centrosymmetric ones having large strains, containing impurities with a large spin-orbit coupling and in ultra-thin ferromagnet where the inversion-symmetry is broken.[17]

Here, we perform a numerical study of the FMR response by means of micromagnetic simulations by using a full micromagnetic framework in which both the SHE and the DMI are implemented. We studied an experimental system similar to the one reported by *Liu et al.*[18] The MTJ stack is made by (CoFeB(1)/MgO(1.2)/CoFeB(4)/Ta(5)/Ru(5) (thicknesses in nm)), milled over a Tantalum (Ta) strip (6000 nm x 1200 nm x 6 nm). We introduce a Cartesian coordinate system where the *x*-axis is positioned along the larger dimension of the Ta strip, the *y*-and *z*-axes are consequently oriented along the other in-plane direction and along the thickness of the Ta strip. The MTJ has an elliptical cross section 180 x 50 nm$^2$ with the larger dimension oriented in the *y*-direction. The ultra-thin CoFeB(1) acts as FL (saturation magnetization $M_s$=1x10$^6$ A/m) and because of the very low thickness, the interfacial perpendicular anisotropy is large enough ($K_u$=7x10$^5$ J/m$^3$ related only to the interfacial anisotropy energy contribution)[19] to impose an out-of-plane easy axis. The CoFeB(4) is the reference layer (RL) and its magnetization is in-plane fixed along the negative *y*-direction. We apply an external magnetic field $H_{ext}$=8 mT in the negative *y*-direction to balance the dipolar field from the RL. In order to analyse the magnetization dynamics, we numerically solve the following non-linear differential equation, which includes the STT and the spin-orbit torque from the SHE:[20]

$$\frac{d\mathbf{m}}{\gamma_0 M_S dt} = -\frac{1}{(1+\alpha^2)}\mathbf{m}\times\mathbf{h_{EFF}} - \frac{\alpha}{(1+\alpha^2)}\mathbf{m}\times\mathbf{m}\times\mathbf{h_{EFF}}$$
$$-\frac{d_J}{(1+\alpha^2)\gamma_0 M_S}\mathbf{m}\times\mathbf{m}\times\boldsymbol{\sigma} + \frac{\alpha d_J}{(1+\alpha^2)\gamma_0 M_S}\mathbf{m}\times\boldsymbol{\sigma}$$
$$+\frac{g}{|e|}\frac{|\mu_B|J_{MTJ}}{\gamma_0 M_s^2 t}g_T(\mathbf{m},\mathbf{m_p})\left[\mathbf{m}\times(\mathbf{m}\times\mathbf{m_p})-q(V)(\mathbf{m}\times\mathbf{m_p})\right] \qquad (1)$$

being **m** and **m$_p$** the magnetizations of the FL and RL respectively. **h$_{EFF}$** is the FL effective field and it contains, as well as the standard magnetic fields, the magnetostatic coupling between the FL and RL and the DMI contribution. The DMI energy density is expressed by:[17]

$$\varepsilon_{DMI} = 2D\left[m_z\nabla\cdot\mathbf{m}-(\mathbf{m}\cdot\nabla)m_z\right] \qquad (2)$$

where, because of the ultra-thin free layer, the magnetization spatial variation along the *z*-direction is neglected $\left(\frac{\partial\mathbf{m}}{\partial z}=0\right)$. *D* is the parameter which takes into account the intensity of the DMI. From the last equation, we can derive the additional term to the effective field related to the DMI: $\mathbf{h_{DMI}} = -\frac{1}{\mu_0 M_S}\frac{\partial\varepsilon_{DMI}}{\partial\mathbf{m}}$. Furthermore, *g* is the Landè factor, $\mu_B$ is the Bohr magneton, $\mu_0$ is the vacuum magnetic permeability, *e* is the electron charge, $\gamma_0$ is the gyromagnetic ratio, $\alpha$ is the Gilbert damping, $M_s$ is the saturation magnetization, *t* is the FL thickness, $J_{MTJ}$ is the current density flowing through the MTJ stack, $g_T(\mathbf{m},\mathbf{m_p})=\frac{2\eta_T}{1+\eta_T^2 \mathbf{m}\cdot\mathbf{m_p}}$ characterizes the angular dependence of the spin-polarization function for the MTJ as computed by Slonczewski,[21, 22] where $\eta_T$ is the polarization efficiency. $q(V)$ is a function which takes into account the squared voltage dependence of the field-like torque up to a maximal value equal to the 25% of the in-plane torque.[23, 24, 25] The coefficient $d_j$ is given by $d_J = \frac{\mu_B \alpha_H}{eM_S t}J_{Ta}$, in which $\alpha_H$ is the spin Hall angle obtained by the ratio between the amplitude of the $J_{SHE}$ and the tantalum current $J_{Ta}$.[26, 27] $\sigma$ is the direction of the $J_{SHE}$ in the Ta strip. The magnetic parameters for the micromagnetic study are: exchange constant $A$=2.0 x $10^{-11}$ J/m, magnetic damping $\alpha$=0.021 and spin-hall angle $\alpha_H$=-0.15.

Firstly, we consider the STT effect (FMR response with no $J_{Ta}$) for a $J_{MTJrf} = J_{MAX}\text{sen}(2\pi f_{rf}t)$ with an amplitude $J_{MAX}$=0.5x$10^6$ A/cm$^2$ and sweeping its frequency $f_{rf}$ from 3.0 GHz to 7.6 GHz. The FMR signal is computed as the difference between the maximum and the minimum value of the oscillating $y$-component of the average magnetization. Two scenarios are investigated: the first one, where the DMI effect is neglected ($D$=0 mJ/m$^2$) and the other one when the DMI is relevant ($D$=-1.2 mJ/m$^2$).[17] Without the DMI contribution, the FMR shows only one peak at 5.8 GHz, (see Fig.1a upper curve). The insets near to the peak of Fig. 1a illustrate the spatial mode distributions (SMDs) for the $y$- and $z$-component of the magnetization at the FMR frequency, as computed with the micromagnetic spectral mapping technique.[28] As can be noted, a central mode is excited for the two magnetization components. By considering the DMI (see lower curve), the FMR response displays two peaks: the first one at a frequency of 5.7 GHz (indicated with 1 in Fig. 1a) and the second one at 6.1 GHz (indicated with 2 in Fig. 1a). This FMR behavior is clearly due to the effect of the DMI, which splits the FMR mode in two, as also observed in the SMDs. In fact, while the SMD of the $y$-component shows a similar central mode, the SMD of the $z$-component displays the generation of two edge modes for the first peak and four edge modes for the second peak. Moreover, observing the time domain plot (not represented here), we note that for the frequency of 5.7 GHz, the TMR is in advance with respect to the injected microwave current; on the contrary, at the frequency of 6.1 GHz, the TMR is lagging behind the $J_{MTJrf}$. The DMI influences only the *z*-component because of the edge non-uniformities induced by the dipolar field.

Fig. 1b represents the FMR responses for a microwave current of $J_{MAX}$=0.1x10$^6$ A/cm$^2$. The FMR frequency increases, either without DMI (top curve), reaching 6.1 GHz, or with DMI (bottom curve), attaining 6.2 GHz and, additionally, both the FMR curves have a single peak. The evidence of only one frequency peak, even considering the DMI, is ascribed to the use of a weak microwave current, which keeps the FMR response in a linear regime. Also in this case, the SMD of the magnetization *z*-component shows an edge mode. In addition, the FMR frequency changes (value and shape) with $J_{MAX}$ because a higher amplitude of the microwave current generates non-linear dynamics.[2, 8] As demonstrated in Ref. [23] (see Figure 3a), this non-uniform regime can be observed by the presence of an asymmetric FMR spectrum.

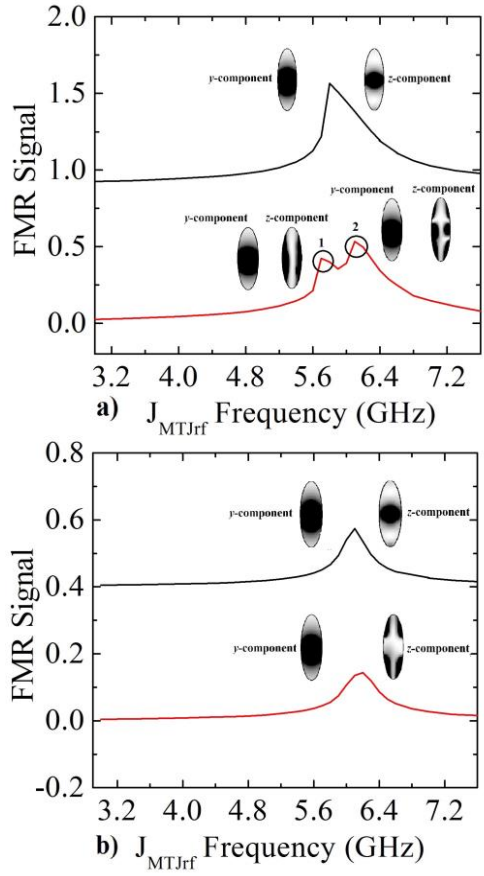

Fig. 1: FMR responses for $J_{Ta}=0$ A/cm$^2$. a) $J_{MAX}=0.5 \times 10^6$ A/cm$^2$ without DMI (top curve) and with DMI (bottom curve); b) $J_{MAX}=0.1 \times 10^6$ A/cm$^2$ with no DMI (upper curve) and with DMI (lower curve). The insets represent the SMDs for the *y*- and *z*- components of the magnetization.

Fig. 2 shows the FMR response when a bias $J_{Ta}=-1.50 \times 10^7$ A/cm$^2$ flows in the Ta strip with (upper curve) and without (lower curve) the DMI contribution. A microwave current of $J_{MAX}=0.5 \times 10^6$ A/cm$^2$ is injected in the MTJ stack. The top curve shows a similar behavior with respect to the corresponding curve without the in-plane current (Fig. 1a); in fact, a main central mode is visible in the SMDs for both *y*- and *z*-component of the magnetization. A similar behavior is also obtained in presence of the DMI: two FMR peaks are visible and the SMDs have a configuration similar to the one previously investigated (see for comparison the SMDs in Fig. 1a lower curve). Hence, the FMR behavior is not affected by a sub-critical $J_{Ta}$, and the DMI effect concerns again the splitting of the main mode.

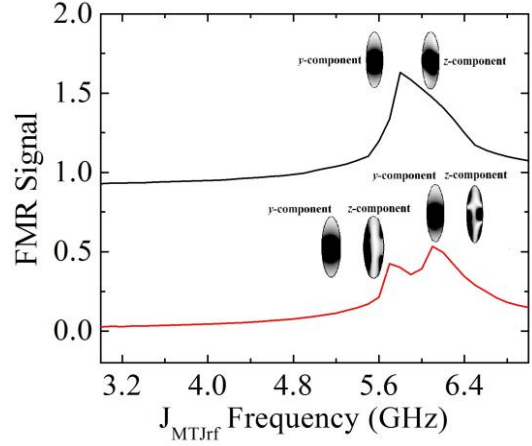

Fig. 2: FMR responses for a sub-critical $J_{Ta}$=-1.50x10$^7$ A/cm$^2$ and a $J_{MAX}$=0.5x10$^6$ A/cm$^2$ without DMI (top curve) and with DMI (bottom curve). The insets represent the SMDs for the *y*- and *z*- components of the magnetization.

The FMR behavior is different when the $J_{Ta}$ is increased. Fig. 3a shows the computed FMR with (upper curve) and without (lower curve) DMI, when both the $J_{Ta}$=-1.40x10$^8$ A/cm$^2$ (this value is very close to the switching one, which leads the FL from out-of-plane to in-plane) and the microwave current with amplitude $J_{MAX}$=0.5x10$^6$ A/cm$^2$ are applied. Without the DMI, the increasing of $J_{Ta}$ does not change the FMR qualitatively, but it induces a reduction of the FMR frequency, from 5.8 GHz (Fig. 1a) to 4.8 GHz. Whereas, taking into account the DMI, a decreasing of the FMR frequency and a low power peak at higher frequency are observed. Furthermore, the SMDs of the lower frequency peak show that the main excited mode is shifted from the central position. Thus, with a $J_{Ta}$ large enough (that means a relevant SHE contribution), the DMI moves the SMD of the main central mode towards the left side of the sample. Changing the sign of *D*, the central mode moves to the right side (not shown). Fig. 3b displays the FMR for $J_{Ta}$=-1.40x10$^8$ A/cm$^2$ and $J_{MAX}$=0.1x10$^6$ A/cm$^2$. In this case, while the FMRs with and without the DMI are very similar (top and bottom curves respectively), a small displacement of the central mode is observed in the resonance frequency SMDs including the DMI.

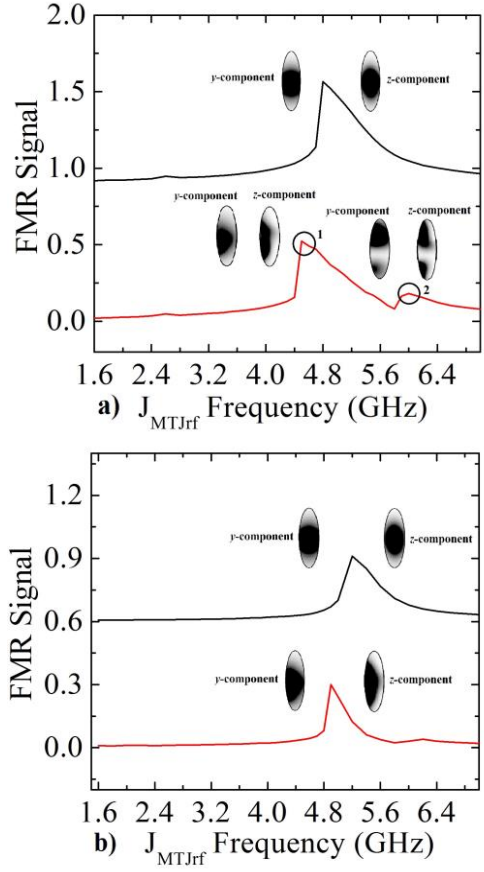

Fig. 3: FMR responses for $J_{Ta}$=-1.40x10$^8$ A/cm$^2$. a) $J_{MAX}$=0.5x10$^6$ A/cm$^2$ without DMI (top curve) and with DMI (bottom curve); b) $J_{MAX}$=0.1x10$^6$ A/cm$^2$ with no DMI (upper curve) and with DMI (lower curve). The insets represent the SMDs for the *y*- and *z*- components of the magnetization.

In summary, the effect of the DMI on the FMR has been analyzed in both cases with and without the in-plane Ta current. We have observed that, regardless of the $J_{Ta}$, the effect of the DMI is to break the symmetry of the main central excited mode. However, the way of the symmetry breaking has been dependent on the Ta current. If $J_{Ta}$ is negligible or it assumes a sub-critical value, we have observed that DMI breaks the symmetry of the main central mode in two (or more, as function of the peak) different edge modes. For larger $J_{Ta}$, the DMI contribution is not so great to divide the main excited mode and its effect is the displacement of the main central mode towards a lateral position only.